\documentclass[utf8]{frontiersFPHY} 

\usepackage{url,hyperref,lineno,microtype,subcaption}
\usepackage[onehalfspacing]{setspace}

\usepackage{amsmath}
\usepackage{bm}

\def\keyFont{\fontsize{8}{11}\helveticabold }
\def\firstAuthorLast{Jose P. Rodriguez} 
\def\Authors{Jose P. Rodriguez\,$^{1,*}$}
\def\Address{$^{1}$Department of Physics and Astronomy, California State University at Los Angeles, Los Angeles,
 California, USA}
\def\corrAuthor{Department of Physics and Astronomy, California State University at Los Angeles, Los Angeles,
 CA 90032, USA}

\def\corrEmail{jrodrig@calstatela.edu}

\begin{document}
\onecolumn
\firstpage{1}

\title[Quantum-Critical Spin-Density Waves in Iron-Selenide High-$T_c$ ...]
{Quantum-Critical Spin-Density Waves in Iron-Selenide High-$T_c$ Superconductors}

\author[\firstAuthorLast ]{\Authors} 
\address{} 
\correspondance{} 

\extraAuth{}

\maketitle

\begin{abstract}

\section{}
Hidden spin-density waves (hSDW) with N\'eel ordering vector $(\pi,\pi)$ have been proposed recently
 as parent groundstates to electron-doped iron-selenide superconductors.
Doping such groundstates can result in visible electron-type Fermi surface pockets and
faint hole-type  Fermi surface pockets at the corner of the folded Brillouin zone.
A Cooper pair instability that alternates in sign between the electron-type and the hole-type Fermi surfaces 
has recently been predicted.  The previous is due to the interaction of electrons and holes 
with hidden spin fluctuations connected with hSDW order that is near a quantum-critical point.  
Quantum criticality is tuned in by increasing the strength of Hund's Rule from the hSDW state.
We find that the exchange of hidden spin fluctuations by electrons/holes in
the critical hSDW state results in asymptotic freedom.  In particular, the strength of spin-flip
interactions becomes weaker and weaker on length scales that are shorter and shorter compared to the
range of hSDW order.  We then argue that string states 
that connect well-separated particle/hole excitations in the hSDW are robust.  
This suggests a picture where the hole degrees of freedom mentioned previously are confined.

\tiny
 \keyFont{ \section{Keywords:} superconductivity, iron superconductors, magnetism, renormalization group,
 asymptotic freedom, quantum criticality, spin fluctuations, spin-density waves}
\end{abstract}

\section{Introduction}
Electron-doped iron-selenide systems show among the highest critical temperatures inside the general class of
iron superconductors. A single layer of iron selenide over the appropriate substrate exhibits a critical
temperature on the order of $50$ K, for example\cite{zhou_13,zhang_14}.
  Unlike iron-pnictide superconductors, however,
electron-doped iron selenide shows no nesting of the Fermi surfaces.
Instead, angle-resolved photoemission spectroscopy (ARPES) finds evidence for
 electron-type Fermi surface pockets at the corner of
 the folded (two-iron) Brillouin zone\cite{qian_11,peng_14,lee_14,zhao_16}.
The hole bands that typically cross the Fermi level at the center of the Brillouin zone
in iron-pnictide superconductors
lie buried below the Fermi level in electron-doped iron selenide.
Furthermore, ARPES and scanning tunneling microscopy (STM) find evidence for an {\it isotropic}
quasi-particle gap that opens at the Fermi level\cite{zhou_13,peng_14,lee_14,zhao_16,fan_15,yan_15}.
It is believed that electron-electron repulsion
is moderately strong in iron superconductors generally\cite{mazin_08,kuroki_08}. 
 The absence of sign changes in
the gap over the Fermi surface that is observed experimentally by ARPES and by STM in
electron-doped iron selenide is therefore very puzzling.

Electron-doped iron selenide also exhibits peculiarities in the nature of its low-energy spin excitations.
Inelastic neutron scattering spectroscopy finds evidence for spin resonances at excitation
energies that lie below the superconducting energy gap\cite{park_11},
but at momenta that lie mid-way between the ``stripe'' spin-density wave (SDW) wavevector $(\pi,0)$
and the N\'eel SDW wavevector $(\pi,\pi)$ \cite{friemel_12,davies_16,ma_17}.
Furthermore, inelastic neutron scattering spectroscopy sees
 a ``diamond'' of spin excitations centered at the wavevector $(\pi,\pi)$
in electron-doped iron selenide\cite{pan_17}.
The ``diamond'' of spin excitations notably lies at energies {\it above} the superconducting gap. 

Upon electron doping,
the author has demonstrated recently that a critical hidden spin-density wave (hSDW) state
with N\'eel wavevector $(\pi,\pi)$ shows electron-type
Fermi surface pockets that are accompanied by vanishingly faint hole-type Fermi surface pockets
at the corner of the folded (two-iron) Brillouin zone\cite{jpr_20b}.
This result is consistent with those of a related local-moment model\cite{jpr_17}.
The author has also demonstrated that a Cooper pair instability exists over the electron and hole Fermi surfaces
that alternates in sign between them.  At half filling,
the critical hSDW shows hidden magnetic order between
$d+ = d_{xz} + i d_{yz}$ and $d-  = d_{xz} - i d_{yz}$ orbitals\cite{jpr_rm_18}.
(See Fig. \ref{sdw_hsdw_states}a.)
Lowering the strength of  Hund's Rule coupling 
tunes the critical hSDW into the hidden order  phase\cite{jpr_ehr_09,jpr_10}.
The author has also demonstrated that the hSDW shows ``rings'' and ``diamonds'' of low-energy spin excitations
around the N\'eel wave vector $(\pi,\pi)$\cite{jpr_rm_18,jpr_20a}.
  Such low-energy spin excitations collapse to zero energy at the critical hSWD.

The electron-doped hSDW described above is a promising candidate groundstate
 for electron-doped iron-selenide high-temperature superconductors.
It is predicted to be an $S^{+-}$-wave superconductor,
with Cooper pairs that alternate  in sign between visible electron Fermi surface pockets
and faint hole Fermi surface pockets at the corner of the folded (two-iron) Brillouin zone\cite{jpr_20b}.
It is controlled, however, by the critical hSDW at half filling. (See Fig. \ref{mft_vs_qcp}b.)
Related critical hSDW states have been studied previously in
the context of copper-oxide high-temperature superconductors\cite{BMS_12}.
(See Fig. \ref{sdw_hsdw_states}c.)
They are notably free of the sign problem in quantum Monte Carlo simulations.
In this paper, we shall continue to scrutinize the critical hSDW at half filling
via an Eliashberg Theory analysis in the particle-hole channel\cite{jpr_rm_18}.
Here, electrons and holes interact with hidden spinwaves that result from long-range hSDW order.
The spinwaves disperse acoustically from the N\'eel wavevector $(\pi,\pi)$.
A renormalization group is discovered that is based on the correction to the interaction vertex.
It predicts asymptotic freedom\cite{politzer_73,gross_wilczek_73,ramond_81,polyakov_87}: 
spin-flip interactions become weaker and weaker
 at shorter and shorter length scales compared to the range of hSDW order.
At the opposite extreme,
we argue that this is related to confining {\it string} states between 
well-separated electron/hole excitations\cite{trugman_88,Shraiman_Siggia_88a,dagotto_94}.
(See Fig. \ref{string_states}.)

\section{Nested Fermi Surfaces and Hidden Spin-Density Wave}
An extended Hubbard model for a single layer of electron-doped iron selenide  is introduced below.
Mean field theory will reveal that it harbors hidden magnetic order\cite{jpr_rm_18}.

\subsection{Extended Hubbard Model}
We retain only the $3d_{xz}/3d_{yz}$ orbitals of the iron atoms
in the following description
of a single layer of heavily electron-doped FeSe. In particular,
let us work in the isotropic basis of orbitals
$d-=(d_{xz}-id_{yz})/{\sqrt 2}$ and $d+=(d_{xz}+id_{yz})/{\sqrt 2}$.
Electrons hop among the $d+$ and $d-$ orbitals between nearest neighbors ($1$)
and next-nearest neighbors ($2$) on the square lattice of iron atoms that make up a single layer of FeSe.
The kinetic energy due to hopping over the square lattice of iron atoms
is then governed by the following Hamiltonian expressed in momentum space:
\begin{equation}
H_{\rm hop} =
\sum_{\bm k}\sum_{\alpha}\sum_s \varepsilon_{\parallel}({\bm k})c_{\alpha,s}^{\dagger}({\bm k}) c_{\alpha,s}({\bm k}) +
\sum_{\bm k}\sum_s [\varepsilon_{\perp}({\bm k}) c_{d+,s}^{\dagger}({\bm k}) c_{{\bar{d-}},s}({\bm k}) + {\rm h.c.}],
\label{hop}
\end{equation}
where
%
\begin{subequations}
\begin{align}
\label{mtrx_lmnt_a}
\varepsilon_{\parallel}({\bm k}) &= -2 t_1^{\parallel} (\cos k_x a + \cos k_y a)
-2 t_2^{\parallel} (\cos k_+ a + \cos k_- a) \\
\label{mtrx_lmnt_b}
\varepsilon_{\perp}({\bm k}) &= -2 t_1^{\perp} (\cos k_x a - \cos k_y a)
-2 t_2^{\perp} (\cos k_+ a - \cos k_- a)
\end{align}
\end{subequations}
%
are intra-orbital and inter-orbital matrix elements,
with  $k_{\pm} = k_x \pm k_y$.
Above also, $c_{\alpha,s}({\bm k})$ and $c_{\alpha,s}^{\dagger}({\bm k})$
denote operators that destroy and create
an electron of spin $s$ in orbital $\alpha$, and with momentum $\hbar {\bm k}$.
The reflection symmetries shown by a single layer of FeSe imply
that the intra-orbital ($\parallel$) and inter-orbital ($\perp$) hopping matrix elements
of the Hamiltonian (\ref{hop})
show $s$-wave and $d$-wave symmetry, respectively\cite{raghu_08,Lee_Wen_08,jpr_mana_pds_14}.
The momentum dependence of the intra-orbital (\ref{mtrx_lmnt_a})
 and inter-orbital (\ref{mtrx_lmnt_b}) matrix elements
$\varepsilon_{\parallel}({\bm k})$ and $\varepsilon_{\perp}({\bm k})$
display those symmetries.
Here,
the nearest neighbor intra-orbital and inter-orbital hopping matrix elements
 $t_1^{\parallel}$ and $t_1^{\perp}$
are real,
the next-nearest neighbor intra-orbital hopping matrix element $t_2^{\parallel}$ is also real,
and   the next-nearest neighbor inter-orbital hopping matrix element $t_2^{\perp}$ is pure imaginary.

The hopping Hamiltonian (\ref{hop}) is easily diagonalized\cite{jpr_rm_18} by
plane waves of $d_{x(\delta)z}$ and $i d_{y(\delta)z}$ orbitals that are rotated
with respect to the principal axis by a phase shift $\delta$
that depends on momentum.
They are created by the corresponding operators:
\begin{eqnarray}
c_s^{\dagger}(2,{\bm k}) = {2}^{-1/2} 
[e^{-i \delta(\bm k)} c_{d-,s}^{\dagger} ({\bm k}) +
e^{+i \delta(\bm k)} c_{d+,s}^{\dagger} ({\bm k})], \nonumber \\
c_s^{\dagger}(1,{\bm k}) = {2}^{-1/2}
[e^{-i \delta(\bm k)} c_{d-,s}^{\dagger} ({\bm k}) -
e^{+i \delta(\bm k)} c_{d+,s}^{\dagger} ({\bm k})].
\label{ck}
\end{eqnarray}
%
The phase shift $\delta({\bm k})$ is set by
$\varepsilon_{\perp}({\bm k}) = |\varepsilon_{\perp}({\bm k})| e^{i 2 \delta({\bm k})}$.
Specifically,
%
\begin{subequations}
\begin{align}
\label{c_2dlt}
\cos\,2\delta({\bm k}) &= {-t_1^{\perp}(\cos\, k_x a - \cos\, k_y a)\over
{\sqrt{t_1^{\perp 2}(\cos\, k_x a - \cos\, k_y a)^2 +
|2 t_2^{\perp}|^2 (\sin\, k_x a)^2 (\sin\, k_y a)^2}}}, \\
\label{s_2dlt}
\sin\,2\delta({\bm k}) &= {2 (t_2^{\perp} / i)(\sin\, k_x a) (\sin\, k_y a)\over
{\sqrt{t_1^{\perp 2}(\cos\, k_x a - \cos\, k_y a)^2 +
|2 t_2^{\perp}|^2 (\sin\, k_x a)^2 (\sin\, k_y a)^2}}}.
\end{align}
\end{subequations}
%
It is notably singular at ${\bm k} = 0$ and ${\bm Q}_{\rm AF}$,
where the matrix element $\varepsilon_{\perp}({\bm k})$ vanishes.
The energy eigenvalues of the bonding ($n=2$) and the anti-bonding ($n=1$) states
 are respectively then given by
$\varepsilon_+({\bm k}) = \varepsilon_{\parallel}({\bm k}) + |\varepsilon_{\perp}({\bm k})|$ and
$\varepsilon_-({\bm k}) = \varepsilon_{\parallel}({\bm k}) - |\varepsilon_{\perp}({\bm k})|$.

\begin{figure}[h!]
\begin{center}
\includegraphics[width=10cm]{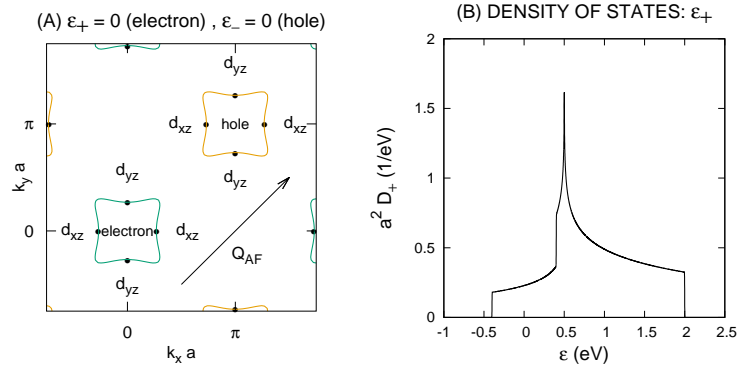}
\end{center}
\caption{(a) Perfectly nested Fermi surfaces at half filling:
$\varepsilon_+({\bm k}) = 0$ and  $\varepsilon_-({\bm k}) = 0$,
with hopping matrix elements
$t_1^{\parallel} = 100$ meV, $t_1^{\perp} = 500$ meV, $t_2^{\parallel} = 0$,
and $t_2^{\perp} = 100\, i$ meV. (See Fig. \ref{bnds_fs1} for the band structure.)
Dirac cones emanate from the dots on each Fermi surface.
(b) Also displayed is the density of states of the bonding
 band, $\varepsilon_+({\bm k})$.}
\label{fs0_dos}
\end{figure}

If we now turn off next-nearest neighbor intra-orbital hopping, $t_2^{\parallel} = 0$,
the above energy bands then satisfy the perfect nesting condition\cite{jpr_rm_18},
\begin{equation}
\varepsilon_{\pm}({\bm k}+{\bm Q}_{\rm AF}) = - \varepsilon_{\mp}({\bm k}).
\label{prfct_nstng}
\end{equation}
Here ${\bm Q}_{\rm AF} = (\pi/a,\pi/a)$ is the checkerboard ordering vector on the square
lattice of iron atoms.
As a result, the Fermi level
 lies at $\epsilon_{\rm F} = 0$ at half filling.
Figure \ref{fs0_dos} shows such
perfectly nested electron-type and hole-type Fermi surfaces for hopping
parameters $t_1^{\parallel} = 100$ meV, $t_1^{\perp} = 500$ meV, $t_2^{\parallel} = 0$
and $t_2^{\perp} = 100\, i$ meV.  It also displays the density of states of the bonding band, $n = 2$.
The step at $0.4$ eV and the sharp peak near $0.5$ eV mark where constant-energy contours in
momentum space experience a change in topology.  
Such transitions will play an important role below.

The principal interactions among the electrons are on-site ones due to Coulomb repulsion.
They have four parts that are included in the following extended Hubbard model
for the interaction Hamiltonian\cite{2orb_Hbbrd}:
\begin{eqnarray}
H_U = & \sum_i &[U_0 \sum_{\alpha} n_{i,\alpha,\uparrow} n_{i,\alpha,\downarrow}
                +J_0 {\bm S}_{i, d-}\cdot {\bm S}_{i, d+} \nonumber \\
               &&+U_0^{\prime} n_{i,d+} n_{i,d-}
                +J_0^{\prime} (c_{i,d+,\uparrow}^{\dagger}c_{i,d+,\downarrow}^{\dagger}
                            c_{i,d-,\downarrow}c_{i,d-,\uparrow}+ {\rm h.c.})].
\label{U}
\end{eqnarray}
Here, $n_{i,\alpha,s} = c_{i,\alpha,s}^{\dagger}c_{i,\alpha,s}$ is the occupation operator
and  ${\bm S}_{i,\alpha} = {1\over 2}\sum_{s,s^{\prime}}
c_{i,\alpha,s}^{\dagger}{\boldsymbol \sigma}_{s,s^{\prime}} c_{i,\alpha,s^{\prime}}$
is the spin operator,
while $n_{i,\alpha} = n_{i,\alpha,\uparrow} + n_{i,\alpha,\downarrow}$.
Above also, $U_0>0$ denotes the intra-orbital on-site Coulomb repulsion energy,
while $U_0^{\prime} > 0$ denotes the inter-orbital one.
Last, $J_0 < 0$ is the Hund's Rule exchange coupling constant, which is ferromagnetic,
while $J_0^{\prime}$ denotes the matrix element for on-site Josephson tunneling between orbitals.

We shall also include  super-exchange interactions
among magnetic moments of neighboring iron atoms
 via the Se atoms\cite{anderson_50,Si&A}:
\begin{eqnarray}
H_{\rm sprx} &=&
\sum_{\langle i,j \rangle}  J_1 ({\bm S}_{i, d-} + {\bm S}_{i, d+})
\cdot ({\bm S}_{j, d-} + {\bm S}_{j, d+}) \nonumber \\
&&+\sum_{\langle\langle  i,j \rangle\rangle} J_2 ({\bm S}_{i, d-} + {\bm S}_{i, d+})
\cdot ({\bm S}_{j, d-} + {\bm S}_{j, d+}).
\label{sprx}
\end{eqnarray}
Above, $J_1$ and $J_2$ are positive
super-exchange coupling constants
over nearest neighbor and next-nearest neighbor iron sites.
Magnetic frustration shall be assumed, henceforth, to be moderate to strong:
$J_2 > 0.5 J_1$,

\begin{figure}[h!]
\begin{center}
\includegraphics[width=10cm]{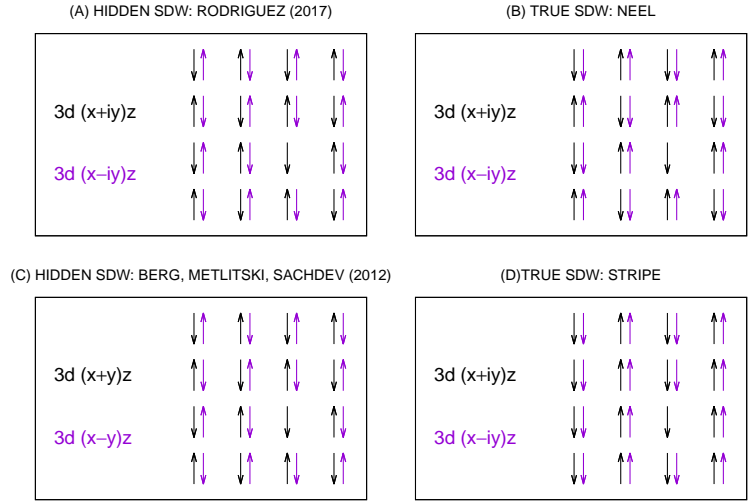}
\end{center}
\caption{Hidden spin-density waves versus true spin-density waves.}
\label{sdw_hsdw_states}
\end{figure}

\subsection{Mean Field Theory}
The perfect nesting of the Fermi surfaces shown by Fig. \ref{fs0_dos}
implies a spin-density wave groundstate for the extended Hubbard model at half filling\cite{jpr_rm_18}.
Possible groundstates are the hSDW state and the true SDW state displayed by
Figs. \ref{sdw_hsdw_states}a and \ref{sdw_hsdw_states}b.
Recent RPA calculations about the hSDW state confirm that
magnetic frustration in $H_{\rm sprx}$ (\ref{sprx}) suppresses
conventional N\'eel order among the iron moments
in favor of hSDW order\cite{jpr_20a}.
A mean field theory for the hSDW state of the extended Hubbard model described above
can be developed along the lines of the mean field theory for the SDW state of the
conventional one-orbital Hubbard model over the square lattice\cite{hirsch_85}.
In particular, assume that the magnetic moment per site per orbital is hidden,
 with spontaneous symmetry breaking along the $z$ axis, and that
it flips sign in a checkerboard fashion across the square lattice of iron atoms:
\begin{equation}
\langle m_{i,\alpha}\rangle = (-1)^{\alpha} e^{i{\bm Q}_{\rm AF}\cdot{\bm r}_i} \langle m_{0,0}\rangle,
\label{ordered_moment}
\end{equation}
where
$\langle m_{i,\alpha}\rangle = {1\over 2}
\langle n_{i,\alpha,\uparrow}\rangle-{1\over 2}\langle n_{i,\alpha,\downarrow}\rangle$.
Here, the $d-$ and $d+$ orbitals are indexed by $\alpha = 0$ and $1$.
It is important to notice that
this hSDW state is notably invariant under rotations of the $3 d_{xz}$ and $3d_{yz}$
orbitals about the $z$ axis.  
It therefore cannot show any tendency towards nematic order\cite{nandi_10,yoshizawa_simayi_12}.
A standard mean-field replacement of the intra-orbital on-site interactions ($U_0$)
 and of the Hund's Rule coupling ($J_0$)
yields the contribution to the electronic energy\cite{jpr_rm_18}
\begin{equation}
-\sum_i \sum_{\alpha} U(\pi) \langle m_{i,\alpha}\rangle (n_{i,\alpha,\uparrow} - n_{i,\alpha,\downarrow})
= - \langle m_{0,0} \rangle  U(\pi)
\sum_i \sum_{\alpha}
(-1)^{\alpha} e^{i{\bm Q}_{\rm AF}\cdot{\bm r}_i}
(n_{i,\alpha,\uparrow} - n_{i,\alpha,\downarrow}) ,
\label{mf_interaction}
\end{equation}
 where
\begin{equation}
U(\pi) = U_0+{1\over 2} J_0.
\label{U(pi)}
\end{equation}
A similar mean-field replacement of the inter-orbital on-site interactions ($U_0^{\prime}$) 
leads entirely to a shift of the chemical potential\cite{jpr_rm_18}.
Also, no net magnetic moment exists on each iron site within mean field theory because of
hidden magnetic  order.  Hence, the super-exchange  Hamiltonian
(\ref{sprx}) makes no contribution within mean field theory as well.
Last, notice that on-site Josephson tunneling between orbitals ($J_0^{\prime}$)
in $H_U$ is suppressed
at strong on-site-orbital repulsion, $U_0$.  We shall henceforth neglect this contribution
to the extended Hubbard model (\ref{U}) on that basis.

Inverting the planewaves (\ref{ck}) to site-orbital space yields
expressions for the corresponding creation operators:
\begin{eqnarray}
c_{i,d-,s}^{\dagger} = {\mathcal N}^{-1/2} \sum_{\bm k} e^{-i{\bm k}\cdot{\bm r}_i}
e^{+i \delta(\bm k)} [c_{s}^{\dagger} (2,{\bm k}) +
 c_{s}^{\dagger} (1,{\bm k})], \nonumber \\
c_{i,d+,s}^{\dagger} = {\mathcal N}^{-1/2} \sum_{\bm k} e^{-i{\bm k}\cdot{\bm r}_i}
e^{-i \delta(\bm k)} [c_{s}^{\dagger} (2,{\bm k}) -
c_{s}^{\dagger} (1,{\bm k})].
\label{ci}
\end{eqnarray}
%
Here ${\mathcal N} = 2 N_{\rm Fe}$ is the number of iron site-orbitals.
Substitution of expressions (\ref{ci}) and their conjugates
into the right-hand side of the mean-field approximation for the interaction energy (\ref{mf_interaction})
ultimately yields a mean-field Hamiltonian of the form
\begin{eqnarray}
H^{(mf)} &= & \sum_s\sum_{\bm k}
 \varepsilon_+({\bm k})[c_s^{\dagger}(2,{\bm k}) c_s(2,{\bm k})-c_s^{\dagger}(1,{\bar{\bm k}}) c_s(1,{\bar{\bm k}})]
  \nonumber \\
&& + \sum_s\sum_{\bm k}[({\rm sgn}\, s) \Delta({\bm k})
c_s^{\dagger}(1,{\bar{\bm k}}) c_s(2,{\bm k})+{\rm h.c.}],
\label{Hmf}
\end{eqnarray}
where ${\bar{\bm k}} = {\bm k}+{\bm Q}_{\rm AF}$.
The gap function is anisotropic, and it depends on momentum as
\begin{equation}
\Delta({\bm k}) = \Delta_0 \sin[2\delta({\bm k})],
\label{gap}
\end{equation}
with
\begin{equation}
\Delta_0 = \langle m_{0,0}\rangle U(\pi).
\label{Delta0}
\end{equation}
Above,
we have shifted  the momentum of the  anti-bonding band ($n=1$)
by ${\bm Q}_{AF}$, and we have subsequently exploited
 the perfect-nesting condition (\ref{prfct_nstng}).
Above, also, 
intra-band scattering has been neglected because it shows no nesting.

\begin{figure}[h!]
\begin{center}
\includegraphics[width=10cm]{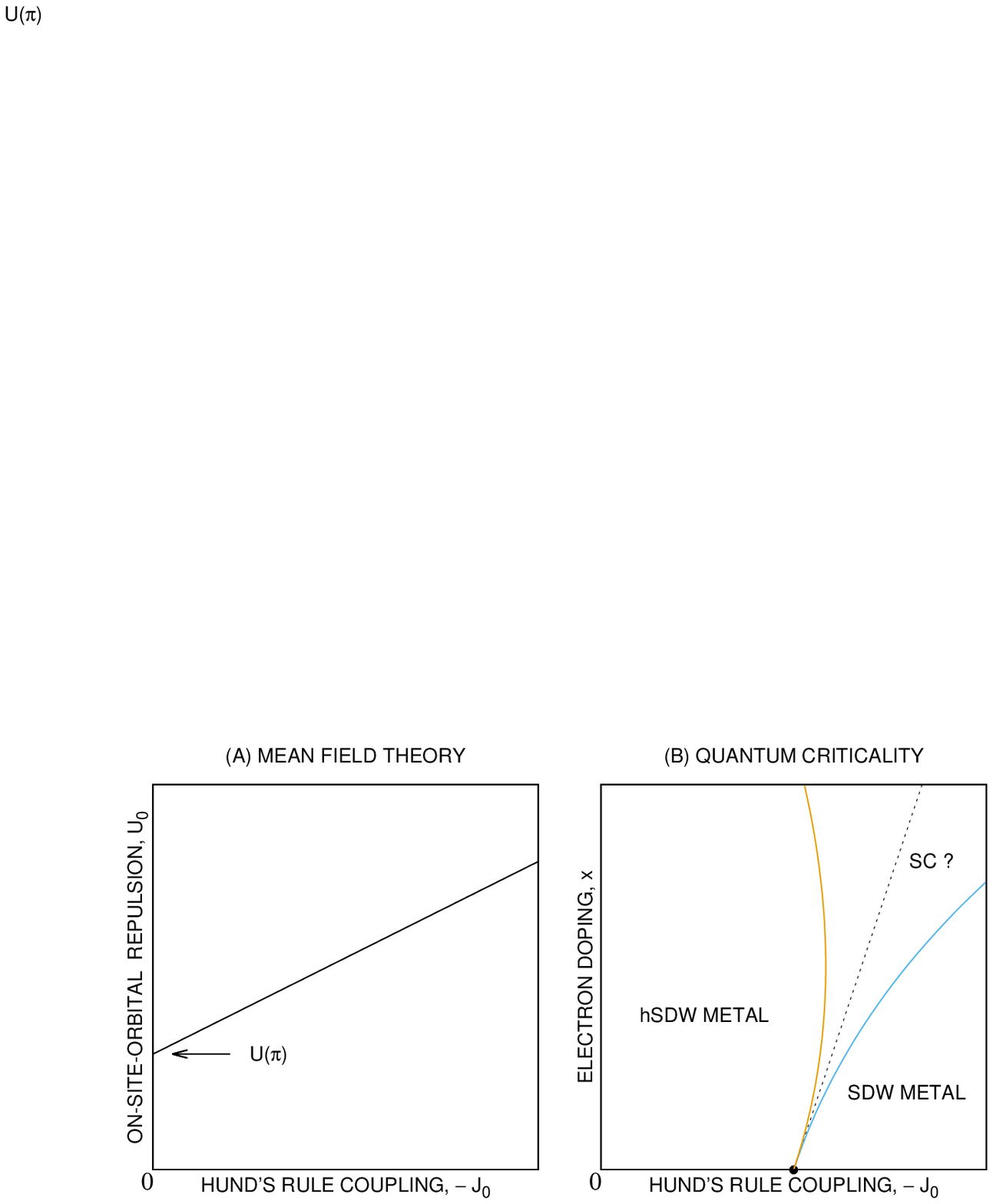}
\end{center}
\caption{(a) Absent quantum critical point (QCP) within mean field theory
versus
(b) QCP, at constant interaction strength,
$U(\pi)$, as Hund's Rule is enforced.}
\label{mft_vs_qcp}
\end{figure}

Inspection of the mean-field Hamiltonian (\ref{Hmf}) yields that it has energy eigenvalues
equal to plus or minus 
$E({\bm k}) = [\varepsilon_+^2({\bm k}) + \Delta^2 ({\bm k})]^{1/2}$.
By (\ref{s_2dlt}), the gap function (\ref{gap}) shows nodes along the principal axes.
Dirac cones therefore emanate from the four points where the Fermi surfaces,
$\varepsilon_+({\bm k}) = 0$ and $\varepsilon_-({\bm k}) = 0$,
cross the principal axes.  These points are displayed by Fig. \ref{fs0_dos}a.
The new one-electron energy spectrum, $\pm E({\bm k})$,
is due to resonant scattering of bonding and anti-bonding planewaves (\ref{ck}) by
the underlying hSDW, which breaks the translational symmetry of the square lattice of iron atoms.  
Amplitudes for the bonding ($+$) and for the anti-bonding ($-$) components
of the new mixed eigenstates are given by the standard {\it coherence factors}.
A {\it gap equation} for the gap maximum (\ref{Delta0}) can be thereby obtained,
which yields the ordered hSDW moment, $\langle m_{0,0}\rangle$,
as a function of the hopping parameters and of $U(\pi)$.
(See ref. \cite{jpr_rm_18}.)
Surprisingly, by (\ref{U(pi)}),
such mean-field solutions remain valid for any pair of
on-site-orbital repulsion and Hund's Rule coupling,
 $U_0$ and $-J_0$, such that $U(\pi)$ remains constant.
Figure \ref{mft_vs_qcp}a displays that degeneracy.

\section{Quantum Field Theory at Criticality}
The hSDW introduced above must clearly become unstable above
a threshold strength in Hund's Rule, $-J_{0c}$.
This fact is not captured by the previous mean field theory.  (See Fig. \ref{mft_vs_qcp}.)
Indeed, the Heisenberg model predicts that the stripe SDW (Fig. \ref{sdw_hsdw_states}d) 
intervenes above a threshold strength in Hund's Rule\cite{jpr_ehr_09,jpr_10}.
Also, a recent RPA calculation of the present extended Hubbard model
is compatible  with the N\'eel SDW  (Fig. \ref{sdw_hsdw_states}b) intervening
 above a threshold Hund's Rule coupling instead \cite{jpr_20a}.
Below, we will include fluctuations of the hSDW order parameter, which
will permit a proper description of the critical hSDW state
that exists at $\Delta_0\rightarrow 0$.

\subsection{Hidden Spinwaves, Interaction with Electrons}
Like in the previous mean field theory, let us start from the ordered hSDW state (\ref{ordered_moment}).
Assume, in particular, a long-range ordered moment
 that alternates in sign between the $d+$ and $d-$ orbitals,
and between nearest neighbor iron sites on the square lattice.
Basic considerations imply Goldstone modes in the transverse directions, $m_x(\pi)$ and $m_y(\pi)$,
at the checkerboard wave number, ${\bm Q}_{\rm AF} = (\pi/a, \pi/a)$,
where ${\bm m}(\pi) = {\bm m}_{d-} - {\bm m}_{d+}$ is the hidden magnetic moment \cite{jpr_rm_18}.
Specifically, the spin-flip propagator,
$iD({\bm q},\omega) =
\langle {1\over{\sqrt{2}}} m^{+}(\pi)
{1\over{\sqrt{2}}} m^{-}(\pi)\rangle |_{{\bm q},\omega}$,
has the hydrodynamic form
\begin{equation}
D({\bm q},\omega) = {(2 s_1)^2\over{\chi_{\perp}}}
[\omega^2 - \omega_b^2({\bm q})]^{-1}
\label{D}
\end{equation}
in the long wave length limit.
Here, $m^{\pm} (\pi) = m_x (\pi) \pm i\, m_y (\pi)$.
The propagator above shows a pole in frequency that disperses
acoustically as  $\omega_b({\bar{\bm q}}) = c_0 |{\bm q}|$,
where ${\bar{\bm q}} = {\bm q} + {\bm Q}_{\rm AF}$,
and where $c_0$ is the hidden-spin-wave velocity.
Above, $s_1$ is the ordered hSDW moment per orbital.
Last, $\chi_{\perp}$ is the transverse spin susceptibility
of the hidden N\'eel state.
This form for the spin-flip propagator (\ref{D}) is required by 
standard antiferromagnetic dynamics\cite{anderson_52,halperin_hohenberg_69,forster_75}.
Recent RPA calculations of the hSDW state reveal such
hidden Goldstone modes that  disperse acoustically\cite{jpr_20a}.

The above hidden spinwaves interact with electrons via
 the intra-orbital Hubbard interaction ($U_0$) and the Hund's Rule coupling ($J_0$) following 
$-\sum_i \sum_{\alpha} U(\pi) {\bm m}_{i,\alpha} \cdot 2{\bm S}_{i,\alpha}$.
This form is suggested by mean field theory (\ref{mf_interaction}),
in which case the super-exchange interaction (\ref{sprx})
makes no contribution in the hSDW state.  Again, on-site Josephson tunneling ($J_0^{\prime}$)
is neglected on the basis that on-site-orbital repulsion $U_0$ is strong enough
to suppress the formation of on-site-orbital singlets.
Keeping only the dominant transverse excitations then yields the interaction
$-\sum_i \sum_{\alpha} U(\pi) (m_{i,\alpha}^+
S_{i,\alpha}^- + m_{i,\alpha}^- S_{i,\alpha}^+$).
Plugging in the transforms (\ref{ci}) between real space and momentum space yields ultimately
the following form of the interaction between hidden spinwaves and electrons\cite{jpr_rm_18}:
\begin{equation}
H_{\rm e-hsw} = + {1\over{\sqrt{2}}}{U(\pi)\over{{a \mathcal N}^{1/2}}}
 \sum_{\bm k} \sum_{{\bm k}^{\prime}}
[m^+(\pi,{\bm q})
{C}_{\downarrow}^{\dagger}({\bm k}^{\prime})
\tau_1
{C}_{\uparrow}({\bm k})
  \sin[\delta({\bm k}) + \delta({\bm k}^{\prime})]
+{\rm h.c.}],
\label{E-hSW}
\end{equation}
where ${\bm q} = {\bm k} - {\bar{\bm k}}^{\prime}$
is the momentum transfer.
Here, we have introduced Nambu-Gorkov spinors\cite{nambu_60,gorkov_58},
\begin{equation}
C_s({\bm k}) =
\left[ {\begin{array}{c}
c_{s}(2,{\bm k}) \\ c_{s}(1,{\bar{\bm k}})
\end{array} } \right],
\label{spinor}
\end{equation}
which explicitly account for perfect nesting of the Fermi surfaces.
Also, $\tau_1$ is the Pauli matrix along the $x$ axis.
Last, intra-band scattering has been neglected because it shows no nesting.

\subsection{Self-Energy Correction and Eliashberg Equations}
We shall now compute the electron propagators within a self-consistent approximation.
They are defined by the $2 \times 2$ matrix
$i G_s({\bm k},\omega) = \int d t_{1,2} e^{i \omega t_{1,2}}
\langle T[C_s({\bm k},t_1) C_s^{\dagger}({\bm k},t_2)]\rangle$,
where $t_{1,2} = t_1 - t_2$, and where $T$ is the time-ordering operator.
Here, 
$C_s({\bm k},t)$ and $C_s^{\dagger}({\bm k},t)$ 
are the time evolutions of the respective destruction and creation operators,
$C_s({\bm k})$ and $C_s^{\dagger}({\bm k})$.
In the absence of interactions, $U(\pi)\rightarrow 0$,
perfect nesting (\ref{prfct_nstng}) yields that
the matrix inverse of the electron propagator is given by
\begin{equation}
G_0^{-1}({\bm k},\omega) =
\omega\, \tau_0 - \varepsilon_+({\bm k})\, \tau_3,
\label{1/G0}
\end{equation}
where $\tau_0$ is the $2 \times 2$ identity matrix,
and where $\tau_3$ is the Pauli matrix along the $z$ axis.
Inspection of the mean-field Hamiltonian (\ref{Hmf})
suggests, on the other hand, the following  form for the matrix inverse of the exact electron propagator:
\begin{equation}
G_s^{-1}({\bm k},\omega) =
Z({\bm k},\omega)  \omega\, \tau_0
- [\varepsilon_+({\bm k})-\nu]\, \tau_3
- ({\rm sgn}\, s) Z({\bm k},\omega) \Delta({\bm k})\,\tau_1.
\label{1/G}
\end{equation}
Here, $Z({\bm k},\omega)$ is the wavefunction renormalization,
$\Delta({\bm k})$
is the quasi-particle gap (\ref{gap}),
and $\nu$ is a relative energy shift of the bands that preserves perfect nesting.
Comparison of (\ref{1/G0}) and (\ref{1/G}) then yields the self-energy correction
\begin{equation}
\Sigma_s({\bm k},\omega) =
[1-Z({\bm k},\omega)] \omega\, \tau_0
-\nu\, \tau_3
+ ({\rm sgn}\, s) Z({\bm k},\omega) \Delta({\bm k})\,\tau_1
\label{Sigma}
\end{equation}
to the band dispersions $\varepsilon_+({\bm k}) \tau_3$.

\begin{figure}[h!]
\begin{center}
\includegraphics[width=10cm]{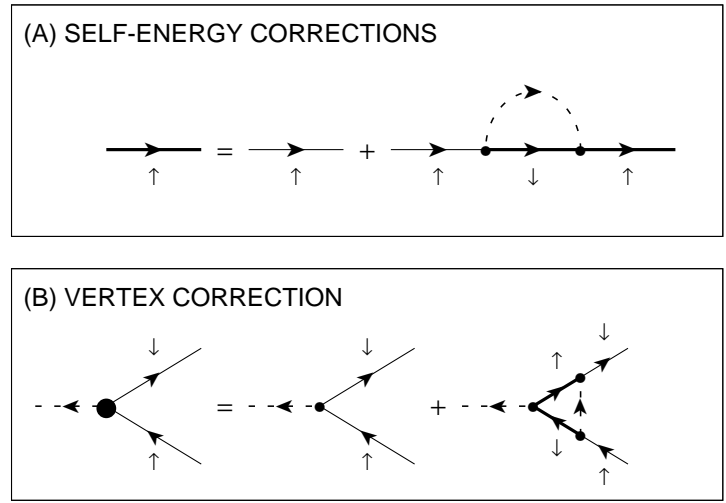}
\end{center}
\caption{Feynman diagrams (a) for the Eliashberg equations
  and (b)  for the second-order vertex correction
to the interaction, Eqs. (\ref{E-hSW}) and (\ref{spinor}).
Thin and thick solid lines represent, respectively, the bare and the ``dressed'' electron propagators,
$G_0$ and $G_s$,
while dashed lines represent the spin-flip propagator, $D$.}
\label{feynman}
\end{figure}

The self-consistent approximation for the electron propagator is displayed by Fig. \ref{feynman}a.
The self-energy correction (\ref{Sigma}) is thereby approximated by
\begin{equation}
\Sigma_s({\bm k},\omega) = 
i \int_{\rm BZ} {d^2 k^{\prime}\over{(2\pi)^2}}  \int_{-\infty}^{+\infty} {d \omega^{\prime} \over{2\pi}}
 {U^2(\pi)\over 2} \sin^2[\delta({\bm k})+\delta({\bm k}^{\prime})]
 D({\bm q},q_0) \tau_1 G_{\bar s}({\bm k}^{\prime},\omega^{\prime})\tau_1,
\label{self-energy}
\end{equation}
with $q_0 =  \omega - \omega^{\prime}$,
and with ${\bm q} = {\bm k} - {\bar{\bm k}}^{\prime}$.
Now write   the electron propagator in terms of components
of Pauli matrices, $\tau_{\mu}$:
 $G = \sum_{\mu = 0}^{3} G^{(\mu)} \tau_{\mu}$.
This yields the corresponding components for the self-energy correction:
\begin{equation}
\Sigma_s^{(\mu)}({\bm k},\omega) =
\, {\rm sgn}_{\mu}(1) \int_{\rm BZ} {d^2 k^{\prime}\over{(2\pi)^2}} 
 {U^2(\pi)\over 2} \sin^2[\delta({\bm k})+\delta({\bm k}^{\prime})]\,
 i \int_{-\infty}^{+\infty} {d \omega^{\prime} \over{2\pi}}
 D({\bm q},q_0)  G_{\bar s}^{(\mu)}({\bm k}^{\prime},\omega^{\prime}).
\label{self-energy_mu}
\end{equation}
Here, we have used the identity satisfied by Pauli matrices,
$\tau_1 \tau_{\mu} \tau_1 = {\rm sgn}_{\mu}(1)\tau_{\mu}$,
where ${\rm sgn}_0(1) = +1 = {\rm sgn}_1(1)$, and
where ${\rm sgn}_2(1) = -1 = {\rm sgn}_3(1)$.
Now recall that the components $G^{(\mu)}$ of the matrix inverse of (\ref{1/G}) are given explicitly by
\begin{eqnarray}
G_s^{(0)} &=& {1\over{2 Z}}
\Biggl({1\over{\omega-E+i\eta}} + {1\over{\omega+E-i\eta}}\Biggr),\nonumber \\
G_s^{(1)} &=& {1\over{2 Z}}
\Biggl({1\over{\omega-E+i\eta}} - {1\over{\omega+E-i\eta}}\Biggr)
{\Delta\over E} ({\rm sgn}\, s), \nonumber \\
G_s^{(2)} &=& 0, \nonumber \\
G_s^{(3)} &=& {1\over{2 Z}}
\Biggl({1\over{\omega-E+i\eta}} - {1\over{\omega+E-i\eta}}\Biggr)
{(\varepsilon_+-\nu)\over Z E},
\label{G}
\end{eqnarray}
with excitation energy
\begin{equation}
E({\bm k},\omega) =
\sqrt{\Biggl[{\varepsilon_+({\bm k})-\nu\over{Z({\bm k},\omega)}}\Biggr]^2
+ \Delta^2({\bm k})}.
\label{E_k}
\end{equation}
Here, $\eta\rightarrow 0+$.
Next, it is useful to write the propagator for hidden spinwaves (\ref{D}) as
\begin{equation}
D({\bm q},\omega) = {(2 s_1)^2\over{\chi_{\perp}}}
{1\over{2\omega_b({\bm q})}}\Biggl[{1\over{\omega - \omega_b({\bm q}) + i\eta}}
-{1\over{\omega + \omega_b({\bm q}) -i\eta}}\Biggr].
\label{d}
\end{equation}
The frequency integrals in expressions (\ref{self-energy_mu}) for the self energies
can be evaluated by going into the complex plane,
and by exploiting
Cauchy's residue theorem in the standard way.
The result
amounts to Brillouin-Wigner second-order perturbation theory\cite{schrieffer_64,scalapino_69}.
In particular,
comparison with the form (\ref{Sigma}) of the self-energy correction yields
the following Eliashberg equations at zero temperature\cite{jpr_rm_18}:
%
\begin{subequations}
\begin{align}
\label{E_eqs_a}
[Z({\bm k},\omega)-1] \omega =&
\int_{\rm BZ} {d^2 k^{\prime}\over{(2\pi)^2}}
 U^2(\pi) {s_1^2\over{\chi_{\perp}}} {\sin^2[\delta({\bm k})+\delta({\bm k}^{\prime})]\over{Z({\bm k}^{\prime},\omega^{\prime})}} \cdot \nonumber\\
& \cdot {1\over{2\omega_b({\bm q})}}
\Biggl[{1\over{\omega_b({\bm q})+E({\bm k}^{\prime})-\omega}}
-{1\over{\omega_b({\bm q})+E({\bm k}^{\prime})+\omega}}\Biggr], \\
\label{E_eqs_b}
-\nu =&
\int_{\rm BZ} {d^2 k^{\prime}\over{(2\pi)^2}}
U^2(\pi) {s_1^2\over{\chi_{\perp}}} {\sin^2[\delta({\bm k})+\delta({\bm k}^{\prime})]\over{Z({\bm k}^{\prime},\omega^{\prime})}}
{\varepsilon_+({\bm k}^{\prime})-\nu\over{Z({\bm k}^{\prime},\omega^{\prime}) E({\bm k}^{\prime})}} \cdot \nonumber\\
& \cdot {1\over{2\omega_b({\bm q})}}
\Biggl[{1\over{\omega_b({\bm q})+E({\bm k}^{\prime})-\omega}}
+{1\over{\omega_b({\bm q})+E({\bm k}^{\prime})+\omega}}\Biggr], \\
\label{E_eqs_c}
 Z({\bm k},\omega) \Delta({\bm k}) =&
 \int_{\rm BZ} {d^2 k^{\prime}\over{(2\pi)^2}}
U^2(\pi) {s_1^2\over{\chi_{\perp}}} {\sin^2[\delta({\bm k})+\delta({\bm k}^{\prime})]\over{Z({\bm k}^{\prime},\omega^{\prime})}}
{\Delta({\bm k}^{\prime})\over{E({\bm k}^{\prime})}} \cdot \nonumber\\
& \cdot {1\over{2\omega_b({\bm q})}}
\Biggl[{1\over{\omega_b({\bm q})+E({\bm k}^{\prime})-\omega}}
+{1\over{\omega_b({\bm q})+E({\bm k}^{\prime})+\omega}}\Biggr],
\end{align}
\end{subequations}
%
where $\omega^{\prime} = E({\bm k}^{\prime})$.  
Below, we shall find a solution to these equations
that describes a quantum-critical hSDW
at $Z({\bm k},\omega) \Delta({\bm k})\rightarrow 0$.

\subsection{Quantum-Critical hSDW, Lifshitz Transition}
The author and Melendrez have shown that a quantum-critical hSDW solution
 of the Eliashberg equations (\ref{E_eqs_a})-(\ref{E_eqs_c})
exists at\cite{jpr_rm_18} $Z({\bm k},\omega) \Delta ({\bm k}) \rightarrow 0$.
Isotropy of the excitation energy $E({\bm k}^{\prime})$ over the Fermi surface is restored in such case.
Like conventional Eliashberg theory for $S$-wave superconductors,
it then becomes convenient to replace the integrals over momentum in (\ref{E_eqs_a}) and (\ref{E_eqs_b})
by integrals over $\varepsilon^{\prime} = \varepsilon_+ ({\bm k}^{\prime})$ and
$\Omega = \omega_b ({\bm q})$.
The previous Eliashberg equations (\ref{E_eqs_a}) and (\ref{E_eqs_b}) thereby reduce to\cite{jpr_rm_18}
%
\begin{subequations}
\begin{align}
\label{2_E_eqs_a}
(Z-1) \omega =&
\int_{-W_{\rm bottom}}^{+W_{\rm top}} d \varepsilon^{\prime}\, Z^{-1}
\int_0^{\omega_{\rm uv}} d\Omega\,  {\epsilon_{\rm E} (\nu)\over{\Omega}} \cdot \nonumber\\
& \cdot {1\over{2}}
\Biggl[{1\over{\Omega+|\varepsilon^{\prime}-\nu|/Z-\omega}}
-{1\over{\Omega+|\varepsilon^{\prime}-\nu|/Z+\omega}}\Biggr], \\
\label{2_E_eqs_b}
\qquad -\nu =&
\int_{-W_{\rm bottom}}^{+W_{\rm top}} d \varepsilon^{\prime}\, Z^{-1}
\int_0^{\omega_{\rm uv}} d\Omega\,  {\epsilon_{\rm E} (\nu)\over{\Omega}} 
{\varepsilon^{\prime}-\nu
\over{|\varepsilon^{\prime}-\nu|}}\cdot \nonumber\\
& \cdot {1\over{2}}
\Biggl[{1\over{\Omega+|\varepsilon^{\prime}-\nu|/Z-\omega}}
+{1\over{\Omega+|\varepsilon^{\prime}-\nu|/Z+\omega}}\Biggr],
\end{align}
\end{subequations}
%
where $\omega_{\rm uv}$ is an ultra-violet frequency cutoff for hidden spinwaves,
and where $[-W_{\rm bottom}, +W_{\rm top}]$ is the range of the energy band $\varepsilon_+({\bm k})$.
(See Fig. \ref{fs0_dos}b.)
Above, we have introduced the Eliashberg energy scale\cite{jpr_rm_18}
\begin{equation}
\epsilon_{\rm E}(\nu) = {1\over{D_+(\nu)}}
\oint_{{\rm FS}_+} {d k_{\parallel}\over{(2\pi)^4}}
 U^2(\pi) {s_1^2\over{\chi_{\perp}}}
{[\sin\, 2\delta({\bm k})]^2\over{c_0 |{\bm v}_+({\bm k})|^2}},
\label{epsilon_E}
\end{equation}
where  FS$_+$ denotes the Fermi surface $\varepsilon_+({\bm k}) = \nu$,
where $D_+(\nu)$ is the density of states of the $n = 2$ bonding band shown by Fig. \ref{fs0_dos}b,
and where $c_0$ is the velocity of hidden spinwaves at ${\bm Q}_{\rm AF}$.
Also, ${\bm v}_+({\bm k}) = \partial \varepsilon_+ / \partial {\bm k}$ is the group velocity.
In obtaining (\ref{epsilon_E}),
the left-hand side of the corresponding Eliashberg equations (\ref{E_eqs_a}) and (\ref{E_eqs_b})
have been averaged over the Fermi surface, $\varepsilon_+ ({\bm k}) = \nu$.
Also, $Z({\bm k}^{\prime},\omega^{\prime})$ has been approximated by $Z({\bm k}^{\prime},\omega)$
on the right-hand side of these equations.

\begin{figure}[h!]
\begin{center}
\includegraphics[width=10cm]{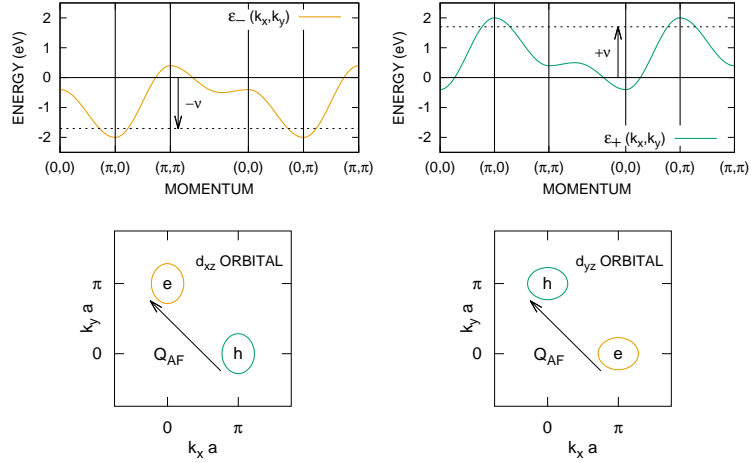}
\end{center}
\caption{Band structure and renormalized Fermi levels at half filling after the Lifshitz transition.
The designated  orbital character of the renormalized Fermi surfaces is only approximate.}
\label{bnds_fs1}
\end{figure}

An analysis of the Eliashberg equations (\ref{2_E_eqs_a}), (\ref{2_E_eqs_b}) and of (\ref{epsilon_E})
reveals a Lifshitz transition as the interaction $U(\pi)$ grows strong,
where the staggered chemical potential $\nu$ approaches the upper edge of the band $\varepsilon_+({\bm k})$.
In particular,
reversing the order of integration in (\ref{2_E_eqs_a}) then yields 
a divergent wave-function renormalization 
at the Fermi level\cite{jpr_rm_18}:
\begin{equation}
Z(\omega)-1 = {\varepsilon_{\rm E}\over{\omega}}, \quad {\rm with} \quad
\varepsilon_{\rm E} = {\pi^2\over 4} \epsilon_{\rm E}.
\label{Z}
\end{equation}
And reversing the order of integration again in (\ref{2_E_eqs_b}) yields the result\cite{jpr_rm_18}
\begin{eqnarray}
\nu =
\epsilon_{\rm E}\, {\rm ln}
\Biggl( {\omega_{\rm uv}\over{\omega_{\rm ir}}} \Biggr)
{\rm ln}
\Biggl( {W\over{\varepsilon_{\rm E}}} \Biggr) 
\label{nu}
\end{eqnarray}
for the staggered chemical potential,
where $\omega_{\rm ir}$ is an infra-red cutoff in the spectrum of hidden spinwaves,
and where $W = W_{\rm bottom} + W_{\rm top}$ is the electronic bandwidth.
Last,
assume that the Lifshitz transition leaves
$\nu$ just below the upper edge of the band $\varepsilon_+({\bm k})$.
Figure \ref{bnds_fs1} depicts
 the later, as well as
 the new Fermi surface pockets that lie at the corner of the folded (two-iron) Brillouin zone.
The Eliashberg energy scale (\ref{epsilon_E}) can be easily estimated in the
case of small circular Fermi surface pockets as $t_1^{\parallel} \rightarrow 0$.
In such case, it is given by\cite{jpr_rm_18}
\begin{equation}
\epsilon_{\rm E} = {1\over 16} \Biggl({x_0\over{2\pi}}\Biggr)^{3/2}
{U^2(\pi)\over{a^2 D_+(\nu)}} {s_1^2\over{a^2 \chi_{\perp}}}
{|t_2^{\perp}|^2\over{(c_0/a) |t_1^{\perp}|^4}}
\label{est_eps_e}
\end{equation}
in the limit of a small concentration of electrons/holes per pocket, $x_0$.
Comparison with (\ref{nu}) yields that $x_0\rightarrow 0$ as
$U(\pi)\rightarrow \infty$.

\subsection{Vertex Corrections, Asymptotic Freedom}
By Fig. \ref{feynman}a,
the previous Eliashberg equations (\ref{E_eqs_a})-(\ref{E_eqs_c})
are obtained from the  summation of ``rainbow'' diagrams
for the electronic self-energy (\ref{self-energy}) \cite{schrieffer_64,scalapino_69}.
Crossing diagrams that lie outside of the previous self-consistent approximation
 are accounted for by including
the correction to the interaction vertex (\ref{E-hSW})  displayed by Fig. \ref{feynman}b.
It reads
\begin{eqnarray}
\gamma^{(2)}(k,k-{\bar q}^{\prime}) & = & i 
\int_{\rm BZ} {d^2 k^{\prime}\over{(2\pi)^2}} \int_{-\infty}^{+\infty}  {d k_0^{\prime}\over{2\pi}}
 \biggl({U(\pi)\over{\sqrt{2}}}\biggr)^3 {\rm sin}[\delta({\bm k})+\delta({\bm k}^{\prime})]
{\rm sin}[\delta({\bm k}^{\prime}-{\bm q}^{\prime})+\delta({\bm k}-{\bm q}^{\prime})] \nonumber \\
&& \cdot {\rm sin}[\delta({\bm k}^{\prime})+\delta({\bm k^{\prime}}-{\bm q}^{\prime})]
D(k-{\bar k}^{\prime}) \tau_1 G_{\downarrow}(k^{\prime}) \tau_1 G_{\uparrow}(k^{\prime}-q^{\prime})\tau_1 ,
\label{gamma_2}
\end{eqnarray}
where ${\bar q}\,^{\prime} = (q_0^{\prime}, {\bm q}^{\prime}  + {\bm Q}_{\rm AF})$.
The integral over frequency can be performed directly, but the calculation is laborious.
Instead, we will exploit a Ward identity to perform that integral\cite{ward_50}.
In particular, by (\ref{1/G}),
taking the derivative of the identity $G_{\bar s} \cdot G_{\bar s}^{-1} = \tau_0$
with respect to $({\rm sgn}\, s) Z \Delta_0$ yields
\begin{equation}
 ({\rm sin}\, 2\delta) G_{\bar s} \tau_1 G_{\bar s} =
-({\rm sgn}\, s) {\partial\, G_{\bar s}\over{\partial (Z \Delta_0)}}. 
\label{Ward_Trick}
\end{equation}
Now recall  (\ref{G}) that $G_{\uparrow} = G_{\downarrow}$ at criticality, $Z \Delta_0\rightarrow 0$.
By (\ref{self-energy}),
substituting (\ref{Ward_Trick}) in for  the relevant factors 
in the integrand on the right-hand side of (\ref{gamma_2})
at $q^{\prime} = 0$
yields the following Ward identity at criticality\cite{ward_50}:
\begin{equation}
\gamma^{(2)}(k,{\bar k}) = -({\rm sgn}\, s){\partial\, \Sigma_s (k) \over{\partial (Z \Delta_0)}}
{U(\pi)\over{\sqrt{2}}} \quad {\rm as} \quad Z \Delta_0\rightarrow 0.
\label{Ward_ID}
\end{equation}
Approaching criticality, $Z \Delta_0 \rightarrow 0$,
comparison of the self-energy correction (\ref{Sigma}) with the gap equation (\ref{E_eqs_c})
 yields a correction to the interaction vertex (\ref{E-hSW}) of the form
\begin{equation}
\gamma^{(2)}(k,{\bar k}^{\prime\prime}) \rightarrow 
\Gamma^{(2)}(k,{\bar k}^{\prime\prime}) {U(\pi)\over{\sqrt{2}}} 
{\rm sin}[\delta({\bm k})+\delta({\bm k}^{\prime\prime})]
\tau_1 
\quad {\rm as} \quad k^{\prime\prime}\rightarrow k ,
\end{equation}
with a vertex correction $\Gamma^{(2)}$ equal to minus the kernel of the gap equation:
$[1+\Gamma^{(2)}] Z \Delta_0 ({\rm sin}\, 2\delta) = 0$.
Expanding the factors of $\sin[\delta({\bm k})+\delta({\bm k}^{\prime})]$
on  the right-hand side of the gap equation
(\ref{E_eqs_c}) by the standard trigonometric identity
reveals the kernel: 
\begin{eqnarray}
-\Gamma^{(2)}(k,{\bar k}) &=& 
{1\over 2}\int_{\rm BZ} {d^2 k^{\prime}\over{(2\pi)^2}}
U^2(\pi) {s_1^2\over{\chi_{\perp}}} {[\sin 2\delta({\bm k}^{\prime})]^2\over{[Z({\bm k}^{\prime},\omega^{\prime})]^2}}
{1\over{E({\bm k}^{\prime})}} \cdot \nonumber\\
&& \cdot {1\over{2\omega_b({\bm q})}}
\Biggl[{1\over{\omega_b({\bm q})+E({\bm k}^{\prime})-\omega}}
+{1\over{\omega_b({\bm q})+E({\bm k}^{\prime})+\omega}}\Biggr].
\label{Gamma_2}
\end{eqnarray}
After substituting 
$Z({\bm k}^{\prime},\omega^{\prime}) = \varepsilon_{\rm E} / E({\bm k}^{\prime})$
in one of the denominators in the integrand above,
then comparison with the second Eliashberg equation (\ref{E_eqs_b}) yields
the vertex correction at criticality,
\begin{equation}
\Gamma^{(2)} \cong
 -{1\over 2} {\nu\over{\varepsilon_{\rm E}}}\,.
\label{Gamma2}
\end{equation}
Here, the factor
$(\varepsilon_+^{\prime}-\nu) / Z^{\prime} E^{\prime}$
on the right-hand side of (\ref{E_eqs_b})
has been set to $-1$
because $Z^{\prime} \Delta^{\prime} = 0$, and because $\nu\rightarrow W_{\rm top}$ as $U(\pi)$ diverges.
Also, the factors of $\sin [\delta({\bm k})+\delta({\bm k}^{\prime})]$
on the right-hand side of (\ref{E_eqs_b})
have been approximated by $\sin[2\delta({\bm k}^{\prime})]$,
which is valid because the factor $1/\omega_b({\bm k}-{\bar{\bm k}}^{\prime})$
diverges as ${\bm k}^{\prime}\rightarrow {\bm k}$.
The vertex correction (\ref{Gamma2}) notably has a negative sign.
Also, inspection of the Eliashberg equation (\ref{nu}) yields the solution
$\nu/\varepsilon_{\rm E} \sim {\rm ln}(\omega_{\rm uv}/\omega_{\rm ir})$.

In the critical hSDW state,
the vertex for the interaction of electrons with hidden spinwaves (\ref{E-hSW})
is multiplied by the factor $\Gamma = 1 + \Gamma^{(2)}$
to lowest non-trivial order,
with a vertex correction given by  (\ref{Gamma2}).
Now call $Y = W/\varepsilon_{\rm E}$.
The second Eliashberg equation (\ref{nu}) can then be re-expressed as
\begin{equation}
\ln\biggl({\omega_{\rm uv}\over{\omega_{\rm ir}}}\biggr) = 
{\nu\over W} {\varepsilon_{\rm E}\over{\epsilon_{\rm E}}} {Y\over{{\rm ln}\, Y}}.
\label{log_omega}
\end{equation}
Here,  $\omega_{\rm uv}$ and $\omega_{\rm ir}$
are the short-range and the long-range cut-offs
in wavelength, respectively, for the spectrum of hidden spinwaves.
First, by (\ref{log_omega}),
it is important to notice that the Eliashberg energy scale $\varepsilon_{\rm E}$ that controls
wavefunction renormalization (\ref{Z}) vanishes logarithmically as $\omega_{\rm ir}\rightarrow 0$.
By Fig. \ref{feynman}b and (\ref{Gamma2}),
repeatedly correcting the interaction vertex (\ref{E-hSW}) to lowest non-trivial order then leads
to the following renormalization group equation 
for the strength of the interaction vertex  as
 a function of the ratio of the short-range ultra-violet scale
to the long-range infra-red scale:
\begin{equation}
{d\Gamma\over{d Y}} = 
-{1\over 2} {\nu\over{W}}
\Gamma^3.
\label{rg_eq}
\end{equation}
%
It has the solution
\begin{equation}
Y = {W\over{\nu}} {1\over{\Gamma^2}} \quad {\rm as} \quad \Gamma\rightarrow 0.
\end{equation}
Second, (\ref{log_omega})
thereby yields the renormalization-group flow
\begin{equation}
\ln\biggl({\omega_{\rm uv}\over{\omega_{\rm ir}}}\biggr) = {\pi^2\over 4} {1\over{\Gamma^2}}
\biggl[{\rm ln}\biggl({W\over{\nu}}{1\over{\Gamma^2}}\biggr)\biggr]^{-1}
\quad {\rm as} \quad \omega_{\rm ir} \rightarrow 0.
\label{rg_flow}
\end{equation}
%
In conclusion, the present quantum field theory is {\it asymptotically free} at criticality:
$\varepsilon_{\rm E}\rightarrow 0$ and $\Gamma\rightarrow 0$
as ${\rm ln} (\omega_{\rm uv} / \omega_{\rm ir})\rightarrow \infty$.
(Cf. refs. \cite{politzer_73}, \cite{gross_wilczek_73}, \cite{ramond_81} and \cite{polyakov_87}.)
If, for example, we impose an $L\times L$ diamond  boundary on the square lattice,
with periodic boundary conditions for the spin-$\uparrow$ electrons,
and with anti-periodic boundary  conditions for the spin-$\downarrow$ electrons,
 then by (\ref{E-hSW}) and (\ref{spinor}),
 $\omega_{\rm ir} /2\pi = c_0/\sqrt{2} L$.
The previous therefore implies that spin-flip interactions become weaker and weaker in the critical hSDW state 
as the size of system approaches the thermodynamic limit.

Last, the previous implies a renormalization-group improved gap equation (\ref{E_eqs_c})
 for the critical hSDW: $\Gamma  Z \Delta_0 \sin (2 \delta) = 0$.
The flow of the renormalization group (\ref{rg_flow}) correctly yields $Z \Delta_0 = 0$
at $\omega_{\rm ir} > 0$ because $\Gamma > 0$ in such case.  
However, $\Gamma \rightarrow 0$ as $\omega_{\rm ir} \rightarrow 0$.
In other words, the kernel of the gap equation is satisfied as the infra-red wavelength diverges.
Such behavior is consisted with assigning the critical hSDW with a quantum critical point.

\section{Discussion and Conclusions}
We have demonstrated above that the quantum-critical hSDW state shows asymptotic freedom:
i.e., interactions become weaker and weaker at shorter and shorter length scales compared to that
for hidden magnetic order, $c_0/\omega_{\rm ir}$. 
Under appropriate boundary conditions,
the latter is of order the size $L$ of the system.
Figure \ref{mft_vs_qcp}b shows a proposed $S^{+-}$ superconducting phase
that is controlled by the critical hSDW state\cite{jpr_20b}.
It has only short-range hSDW order\cite{CHN_88,jpr_90}.
Could the previous renormalization group for asymptotic freedom remain valid there,
but with the infra-red length scale $c_0/\omega_{\rm ir}$ identified with the
finite correlation length for hSDW order instead?
If we  assume this to be true,
then by analogy with the theory of  quantum chromodynamics for
the strong nuclear force\cite{politzer_73,gross_wilczek_73,ramond_81,polyakov_87},
it suggests that the interaction between an electron and a hole in a
quantum-disordered groundstate\cite{CHN_88,jpr_90}
 that is close by to the quantum-critical hSDW state (Fig. \ref{mft_vs_qcp}b)
is due to a  confining string between them.
This should occur at separations that are larger than a suitable scale,
perhaps equal to the correlation length for hSDW order.
Indeed, string states are believed to play a role in the dynamics of the conventional
single-orbital Hubbard model over the square lattice,
in the context of copper-oxide high-temperature superconductors\cite{trugman_88,Shraiman_Siggia_88a}.
It is debated whether confining strings are completely erased by spin-flip interactions,
or whether they persist to some degree\cite{dagotto_94}.
We shall now argue that erasing string states by spin-flips is not likely in states with hSDW order.

\begin{figure}[h!]
\begin{center}
\includegraphics[width=10cm]{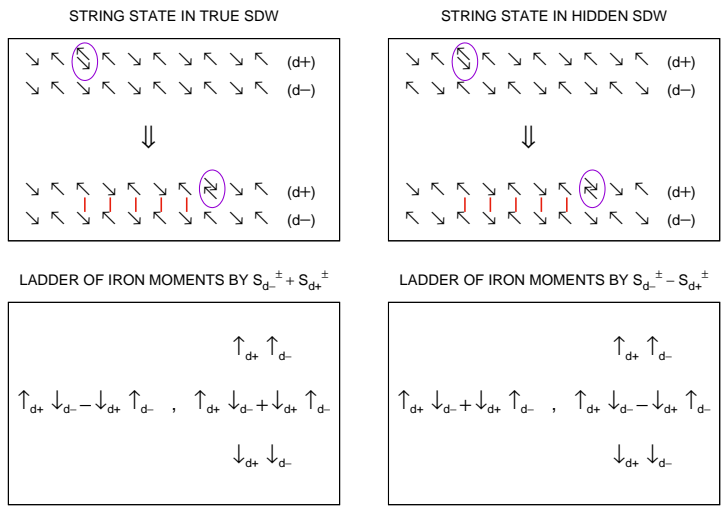}
\end{center}
\caption{String states in the true SDW and in the hSDW.
Also shown are the local irreducible Hilbert spaces that are generated by 
$S_{d+}^{\pm} + S_{d-}^{\pm}$, 
as well as the local irreducible  Hilbert spaces that are generated by
$S_{d+}^{\pm} - S_{d-}^{\pm}$.}
\label{string_states}
\end{figure}

Figure \ref{string_states} shows how the propagation of a spin singlet along a principal axis ($t_1^{\parallel}$)
leaves a string of spins that are out of registry with the
spin on the opposing orbital per iron atom for SDW and for  hSDW order.
Notice that strict long-range order is not essential! 
 In the case of the true SDW state on the left-hand side, application of the raising and lowering
operators for the net spin at a site, $S_{d+,i}^{\pm} + S_{d-,i}^{\pm}$, can raise or lower
the $S_z = 0$ local moments along the string in order to erase it. 
The string state for the hSDW state shown on the right-hand side
of Fig. \ref{string_states} is not the same, however. It is not possible to erase the
string in this case by  application of raising and lowering operators
for the hidden spin at a site, $S_{d+,i}^{\pm} - S_{d-,i}^{\pm}$, 
nor is it possible to erase the string by application of the previous raising and lowering operators
for the net moment. 
The  previous becomes evident by identifying  the ladders of spin states per iron atom that 
such raising and lowering operators live on.  
They are shown on the two bottom panels of Fig. \ref{string_states}.
Repeated applications of $S_{d+}^{\pm} + S_{d-}^{\pm}$ and/or of $S_{d+}^{\pm} - S_{d-}^{\pm}$
on the spin $S_z = \pm 1$ moments in the string (red) of the hSDW
result in either the opposite moment or in the null state.
We therefore conclude that it is unlikely that the string for the hSDW shown in the top-right panel 
of Fig. \ref{string_states} can be erased by spin-flip interactions (\ref{E-hSW}).

In summary, 
we have shown that a quantum-critical hSDW 
that is a candidate parent groundstate for electron-doped iron-selenide superconductors 
shows asymptotic freedom: spin-fluctuation interactions grow weaker and weaker at {\it relatively}
shorter and shorter length scales compared to the size of the system.
The critical hSDW state is tuned in by increasing the strength of Hund's Rule
until hidden magnetic order is no longer stable\cite{jpr_ehr_09,jpr_10}.
We have suggested that such asymptotic freedom 
is a symptom of confining string states
that appear when a single electron propagates in space and time in the presence of hSDW order.  
The string picture is therefore consistent with the vanishingly small weight 
of quasiparticles at the Fermi level at half filling: 
$1/Z(\omega) = \omega/\varepsilon_{\rm E}$
as $\omega\rightarrow 0$.
In addition,
the author has recently shown that electron-doped hSDW states near the critical one studied here
show an $S^{+-}$ Cooper pair instability at the Fermi surface\cite{jpr_20b}.
In particular,
the pair wavefunction alternates in sign between the electron-type and hole-type Fermi surfaces
shown in Fig. \ref{bnds_fs1}.  However, while the quasi-particle weight  of the holes
remains vanishingly small at the Fermi level, the quasi-particle weight of the electrons
can become appreciable at the Fermi level at sufficient electron doping.
Such a dichotomy in the spectral weights between electrons and holes is confirmed by the
spectrum of electron-doped hSDW states within a related local-moment model\cite{jpr_17}.
Within the ``quark'' picture advocated above,
the holes at the hole Fermi surface remain confined and unobservable,
while the electrons at the electron Fermi surface are deconfined and observable.
This suggests searching for the traces of
faint hole bands at the corner of the folded (two-iron) Brillouin zone
in electron-doped iron selenide.

\section*{Conflict of Interest Statement}
The author declares that the research was conducted in the absence of any commercial or financial relationships that could be construed as a potential conflict of interest.

\section*{Funding}
This work was supported in part by the US Air Force
Office of Scientific Research under grant no. FA9550-17-1-0312.

\section*{Acknowledgments}
The author thanks Alex Kass, Ronald Melendrez, Geovani Montoya, and Tong Wang for helpful discussions.

\end{document}